# The origin of anticorrelation for photon bunching on a beam splitter

Byoung S. Ham

Center for Photon Information Processing, School of Electrical Engineering and Computer Science, Gwangju Institute of Science and Technology,
123 Chumdangwagi-ro, Buk-gu, Gwangju 61005, S. Korea
(Dec. 05, 2019)

The Copenhagen interpretation has been long-lasted, whose core concepts are in the Heisenberg's uncertainty principle and nonlocal correlation of EPR. The second-order anticorrelation on a beam splitter represents these phenomena where it cannot be achieved classically. Here, the anticorrelation of nonclassicality on a beam splitter is interpreted in a purely coherence manner. Unlike a common belief in a particle nature of photons, the anticorrelation roots in pure wave nature of coherence optics, where quantum superposition between two input fields plays a key role. This interpretation may intrigue a fundamental question of what nonclassicality should be and pave a road to coherence-based quantum information.

In general, Copenhagen interpretation focuses on the contradictory concept of nonlocal correlation represented by EPR[1]. This nonlocal correlation between two remote objects beyond the physical reality (or local realism) has become the foundation of quantum information processing[2]. The major benchmark for EPR is in the Bell's inequality[3] in a mathematical form or CHSH inequality[4] in its physical version, where the inequality is violated only by quantum mechanics with nonclassical nature. However, the nonlocal correlation of EPR is based on coherence proved by destructive quantum interference-based anticorrelation on a beam splitter (BS)[5], where the nonlocal objects are strictly phase-dependent each other[5,6]. Thus, the definition of classicality in the Copenhagen interpretation must be confined to incoherence optics, representing independent and individual objects with no phase relation[3-5]. This viewpoint makes the present paper unique from others and enriches the conceptual foundation of EPR as well as quantum information.

In both quantum[2] and classical[7] information processing, quantum superposition plays a key role. A typical example of quantum superposition is Young's double-slit experiment effective for both coherent fields (wave nature)[8] and single photons (particle nature)[9-11], resulting in the first-order correlation, $g^{(1)}$. Regardless of the input characteristics, however, the resulting fringe is due to coherence nature between two paths. Thus, the Young's double slit can simply be replaced by a 50:50 nonpolarization BS. The BS-based Young's double slit experiment has also been performed in a Mach-Zehnder interferometer (MZI) with one input[11]. If two input fields are applied to the BS, then the second-order correlation $g^{(2)}$ results in, where each output is a superposed form of both inputs[5]. The sub-Poisson photon statistics of anti-correlation in $g^{(2)}$ indicates that the input photons are anti-bunched nonclassical particles[10]. Here, we discuss that the anticorrelation is a direct result of coherence optics between two input photons with a particular phase relationship. Furthermore, the anticorrelation on a BS can also be achieved in a MZI scheme, and its output superposition results in an entanglement superposition.

The behavior of $g^{(2)}$ relies on the input fields' characteristics[12]: super-Poisson (or $g^{(2)} > 1$) for thermal or chaotic lights; Poisson (or $g^{(2)} = 1$) for coherent lights; sub-Poisson (or $g^{(2)} < 1$) for anti-bunched photons. For $g^{(2)} > 1$, Hanbury Brown and Twiss (HBT) firstly applied it to a high-resolution spectroscopy of distant stars in 1956[13], where the enhanced effect is due to intensity correlation added to the incoherence background ($g^{(2)} = 1$). The opposite case of $g^{(2)} < 1$ was firstly observed by Hong-Ou-Mandel (HOM) using entangled photon pairs generated by a spontaneous parametric down conversion (SPDC) process in 1987[5]. This weird phenomenon of perfect anticorrelation is due to photon bunching into one out of two output paths. On the contrary, coherent input lights is known to follow Poisson statistics of $g^{(2)} = 1$, limiting the HOM dip.

In quantum information, the photon bunching phenomenon on a BS has been intensively studied using anitbunched single photons for basic understanding of nonclassical physics[14-22]. According to Heisenberg's uncertainty principle, a single photon is described as a fixed energy with no particular phase comparable to unlocalized election in the Bohr's atomic model: Copenhagen interpretation. The photon bunching in HOM experiments by single photons seems to be due to incoherence optics at a glance because of its random phase, but it is not, where the lower bound of $g^{(2)}$ by incoherent lights is at most 1/2 (see the Supplementary information A)[19,21,22]. Although a lot of studies have been done for the BS-based $g^{(2)}$ correlation, it is not yet clearly understood what causes the anticorrelation, $g^{(2)}(\tau = 0)=0$. Here, the photon bunching observed in HOM experiments is newly interpreted as a special case of coherence optics, resulting in sub-Poisson statistics of nonclassical nature. This seemingly conflicting result with conventional photon statistics intrigues a fundamental question of what nonclassicality should be. The answer to this question may lead us to better understanding of quantum optics and open a door to quantum superposition-based applications such as unconditionally secured classical key distribution[23] and superposition-enhanced machine learning[24].

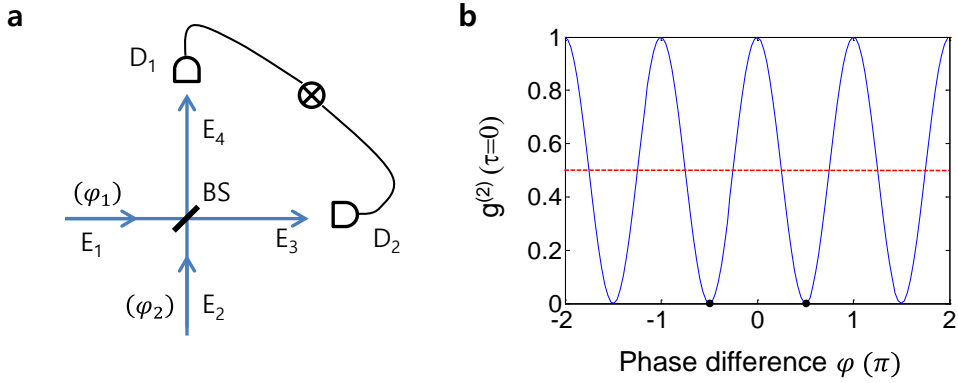

**Figure 1| Intensity correlation. a,** A schematic the second-order interference detection. **b,** Numerical simulation of $g^{(2)}(\tau = 0)$ when $E_1$ and $E_2$ are coherent: Red-dotted line is the lower bound of incoherence optics. BS, beam splitter; $E_i$, i[th] photon field; $D_i$, i[th] photon detector. $\varphi = \varphi_1 - \varphi_2$. Two dots indicate $\varphi = \pm\frac{\pi}{2}$.

The BS matrix, [BS], has been clearly analyzed for the split output fields ($E_3$ and $E_4$) with respect to the input fields ($E_1$ and $E_2$) in 1980[25]. For each input field, [BS] is described by[25]:

$$[BS] = \frac{1}{\sqrt{2}}\begin{bmatrix} 1 & i \\ i & 1 \end{bmatrix}, \qquad (1)$$

where the imaginary number *i* stands for a π/2 phase shift between two coherent outputs. Thus, the output fields in Fig. 1a can be described in a simple matrix form:

$$\begin{bmatrix} E_3 \\ E_4 \end{bmatrix} = [BS]\begin{bmatrix} E_1 \\ E_2 \end{bmatrix}. \qquad (2)$$

To study the second-order correlation in a classical regime of coherence optics, the inputs are set to be traveling light fields with different wavelengths and phases for generality: $E_1 = E_0 e^{i(k_1 r - w_1 t + \varphi_1)}$; $E_2 = E_0 e^{i(k_2 r - w_2 t + \varphi_2)}$, where $k_i$, $w_i$, $\varphi_i$ are wave vector, angular frequency, and initial phase of each field, respectively. The coincidence detection P for two outputs ($E_3$ and $E_4$) represents for the second-order correlation function, $g^{(2)}$:

$$P = \frac{1}{\Delta T}\int_0^{\Delta T} I_1 I_2 dt, \quad (3\text{-}1)$$

$$g^{(2)} = \frac{\langle E_3 E_3^* E_4 E_4^* \rangle}{(\langle E_3 E_3^* \rangle \langle E_3 E_3^* \rangle)}, \quad (3\text{-}2)$$

where $I_1 = E_1 E_1^*$, $I_2 = E_2 E_2^*$, and $\tau$ is the temporal delay between two outputs at detectors. Here, it should be noted that the anticorrelation is satisfied by coincident arrivals of two input photons on a BS to satisfy destructive quantum interference. Thus the decay $\tau$ is confined to the output photons. The two split outputs ($E_3$ and $E_4$) from one input ($E_1$ or $E_2$) by a BS are automatically coherent each other regardless of the bandwidth, photon numbers, and phase fluctuation of the input field. Even for a single photon with random phase, the split outputs (in a form of superposition) are coherent each other. This is the physical origin of self-interference for a single photon (or wave)-based Young's double-slit experiment. Thus, the BS physics is consistent regardless of the photon characteristics whether it is a single photon or waves. The HBT is fulfilled for a single input of chaotic lights satisfying the BS matrix (see the Supplementary Material B). The HOM dip is for two input photons as described in Fig. 1a satisfying the BS coherence optics, too (discussed in Analysis).

Based on two independent input fields of $E_1$ and $E_2$ in Fig. 1a, each output is described as coherent superposition of the inputs from equation (2):

$$E_3 = \frac{1}{\sqrt{2}}(E_1 + iE_2),$$

$$= \frac{1}{\sqrt{2}} E_0 \left( e^{i(k_1 r - w_1 t + \varphi_1)} + i e^{i(k_2 r - w_2 t + \varphi_2)} \right), \quad (4)$$

$$E_4 = \frac{1}{\sqrt{2}}(iE_1 + E_2)$$

$$= \frac{1}{\sqrt{2}} E_0 \left( i e^{i(k_1 r - w_1 t + \varphi_1)} + e^{i(k_2 r - w_2 t + \varphi_2)} \right). \quad (5)$$

According to the definition of the second-order correlation $g^{(2)}$, the intensity-normalized coincidence detection $P' (= g^{(2)}(\tau = 0))$ becomes (see the Supplementary information A):

$$P' = \frac{\langle I_3 I_4 \rangle}{\langle I_3 \rangle \langle I_4 \rangle} = \frac{1}{2}\langle 1 + cos2(\Delta + \varphi) \rangle, \quad (6)$$

where $\Delta = (k_1 - k_2)r - (w_1 - w_2)t$. The relative phase, $\varphi = \varphi_1 - \varphi_2$, is thus fixed regardless of time variation. From equation (6), the solution for the anti-correlation ($P' = 0$) is obtained as:

$$\Delta + \varphi = \pm\left(n - \frac{1}{2}\right)\pi, \quad n=1, 2, 3 \ldots \quad (7)$$

Because lights' path lengths are fixed for coincidence detection in Fig. 1a, the frequency difference-caused phase fluctuation in $\Delta$ dominates in $g^{(2)}$ if two inputs are not degenerate.

Now, we analyze equation (6) for two different categories determined by the input field's phase relationship: Coherence vs. Incoherence. We assume that the input fields are monochromatic for simplicity. The delay time $\tau$ between two outputs at both detectors is set to be zero to satisfy the coincidence detection. Furthermore, a typical HOM dip is analyzed for the proof of concept.

(i) Coherence optics for $g^{(2)}$

For the same wavelength input fields of E$_1$ and E$_2$ ($\Delta = 0$; $\varphi \neq 0$) in Fig. 1a, equation (6) results in $\varphi$−dependent $g^{(2)}$. In this case, two input fields cannot be discernible by the BS or detectors regardless of $\varphi$, satisfying the indistinguishability condition in HOM experiments. This indistinguishability originates in the Young's double-slit experiment using single photons[11]. Here, the input field's bandwidth determines the coherence time of $g^{(2)}$ [5]. Figure 1b shows numerical calculations of the second-order correlation $g^{(2)}$ as a function of $\varphi$ for the coincidence detection ($\tau = 0$). Such $g^{(2)}$ modulation in HOM experiments has already been observed with the input phase (path) control, even though the missing parameter of equation (7) has not been understood, yet[6]. If two input fields are in phase ($\varphi = 0$) or out of phase ($\varphi = \pi$), then Poisson statistics of coherence feature is satisfied for $g^{(2)} = 1$. If there is a $\pm \pi/2$ phase shift between two input fields, then perfect anticorrelation of nonclassical feature results in: $g^{(2)}(\tau = 0) = 0$ (see the dots). This anticorrelation for $\varphi_n = \pm \left(n - \frac{1}{2}\right)\pi$ obviously conflicts with conventional understanding of $g^{(2)}$ correlation based on particle nature[5]. The 'n' in $\varphi_n$ denotes basis in Hilbert space (discussed later). The interesting result is that equation (7) applies only for the inputs before the BS: Depending on the sign selection of $\varphi_n$ in equation (7), the output channel is deterministic for photon bunching (see the Supplementary information A). Combining the symmetric phase pair in $\varphi_n$, e.g., $\varphi_1 = \pm \frac{\pi}{2}$, results in no which-way information in the output paths via superposition of anti-correlation bases. In a brief conclusion, the physical origin of anti-correlation on a BS is not simply due to the single photon's nonclassical nature but due to the destructive quantum interference between two input modes at a specific phase relation satisfying equation (7). The missing parameter $\varphi_n$ in the input mode is the most important discovery in the present study, where it has never been discussed before. Moreover, two single photons used for HOM experiments are pre-determined for a fixed phase relation (discussed later).

(ii) Incoherence optics for $g^{(2)}$

If two input fields E$_1$ and E$_2$ in Fig. 1a are independent each other with random phase fluctuations, the parameter $\Delta$ in equation (6) plays a key role, where the time average of its cosine function becomes zero, resulting in $P' = 1/2$, regardless of $\varphi$: see the red-dotted line in Fig. 1b. Although this value ($P' = 1/2$) indicates the sub-Poisson photon statistics, it is actually the upper bound of classical physics of incoherence optics[19]. However, the violation of the upper bound for $g^{(2)}(\tau = 0) < 0.5$ has also been observed in nondegenerate HOM experiments[22,26-28]. This unexpected observation is not to violate incoherence optics but to prove coherence optics of HOM experiments via phase matching as well as quantum beating phenomena between two nondegenerate SPDC photons (analyzed in Fig. 2).

(iii) A HOM dip

To understand $g^{(2)}$ violation in incoherence optics[27,28], let's look at a typical SPDC process for a HOM dip whether the down-converted photon pairs are degenerate or nondegenerate. Figure 2 shows the result of equation (6) satisfying equation (7) for randomly detuned photon pairs. Figure 2a is for the SPDC generated photons' spectral distribution in a Gaussian function. The unit of $\delta_j$ is GHz but meaningful with respect to the time $\tau$. Due to the wide bandwidth of SPDC-based photon pairs, the HOM dip should satisfy the nondegenerate case of incoherence optics regardless of the pumping method[5,6,14-22,26-28]. However, each down-converted photon pair is always phase matched via $\chi^{(2)}$ SPDC nonlinear process at τ=0: $\boldsymbol{k}_P = \boldsymbol{k}_S + \boldsymbol{k}_I$; $\omega_P = \omega_S + \omega_I$; $\boldsymbol{k}_j$ and $\omega_j$ are respectively a wave vector and angular frequency of the photon j; The subscript of P, S, and I stand for the pump, signal, and idler photons, respectively. Due to the double spontaneous emission decays by $\chi^{(2)}$, the phase $\varphi$ between each photon pair should be $\pi/2$ to compensate the pump field-excited $\pi$ phase shift, inherently satisfying equation (7). In HOM experiments, the average number of photons counted by a photon detector is far less than a million per second. Considering the $g^{(2)}$ processing and detection time is in the order of nanoseconds, there is no more than one photon contributed to the $g^{(2)}$ value per event. Furthermore, each event contributes equally to $g^{(2)}$ value regardless of their detuning $\delta_j$ at the coincidence time[28]. In other words, the initially given phase to individual SPDC photons does not matter to the $g^{(2)}$ value. What matters is $\delta_j$-dependent frequency detuning in $\tau$ as shown in Fig. 2b. If there is more than one photons per event, then the $g^{(2)}$ value should be also deteriorated due to $\delta_j$-caused interference.

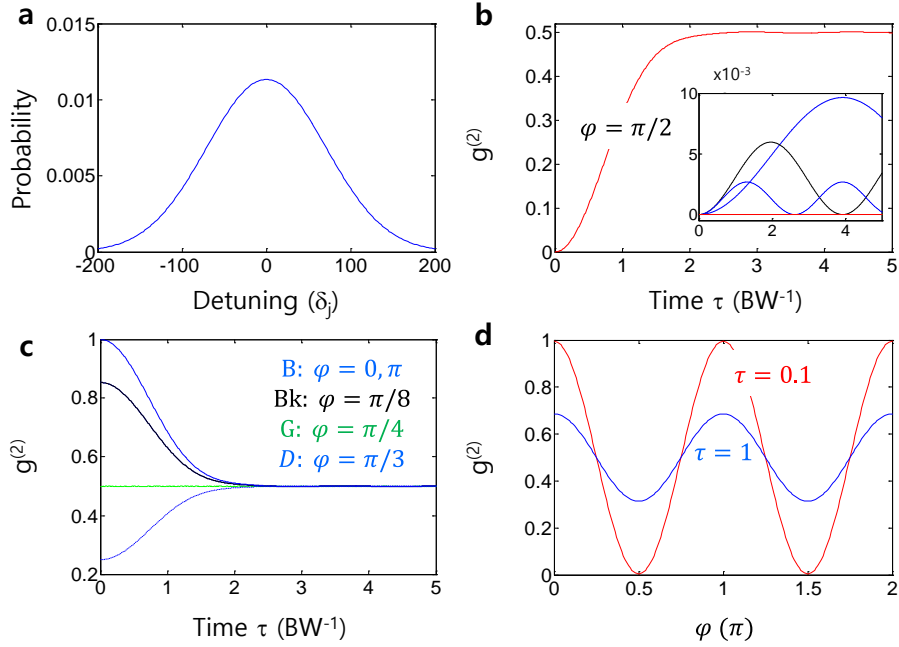

**Figure 2| Numerical analyses for a HOM dip. a,** Photon bandwidth in SPDC-HOM. BW=100 GHz. **b,** Sum of individual photons randomly distributed in **a** for $\varphi = \pi/2$. The inset shows individual signals: R, $\delta_j = 0$; B, $\delta_j = 40$; Bk, $\delta_j = 80$; Dotted, $\delta_j = 120$ GHz. **c,** For different $\varphi$ in **b**. **d,** The $\varphi$-dependent $g^{(2)}$ for different $\tau$. The calculations are based on equation (6). whose detuning step is 2 GHz. The $\tau$ affects only the frequency difference in $\Delta$ of equation (7).

Figure 2b shows the average $g^{(2)}$ for the $\delta_j$-dependent 201 events in Fig. 2a satisfying equation (7) for the anticorrelation condition of $\varphi = \pi/2$. The signal decay in Fig. 2b represents the decoherence due to the $\delta_j$ effect on bandwidth (BW) with the delay $\tau$, where the spatial time delay to detectors is another matter governed by $k_j$ vectors. Even beyond the coherence time, the overall $g^{(2)}$ converges to the upper bound of incoherence optics as expected. This is because that the actual time delay between two output photons is not considered in Fig. 2, because the $\varphi$-dependent $g^{(2)}$ oscillate period is much shorter than the HOM spatial bandwidth[5,6]. The inset of Fig. 2b shows individual $\delta_j$-dependent $g^{(2)}$ evolutions, where the oscillation period relies on $\delta_j$. If equation (7) is violated, then $g^{(2)}$ at $\tau$=0 shows a huge fluctuation depending on $\varphi$ as shown in Fig. 2c. Figure 2d is for $g^{(2)}$ oscillation as a function of the difference phase $\varphi$ (see Fig. 1)[6]. Here, it should be noted that equation (6) applies for both degenerate[5,6] and nondegenerate[27,28] SPDC processes. Due to the wide bandwidth of SPDC, the degenerate case has no practical meaning as discussed above. For the nondegenerate case, a beating signal between two different center-frequencies is just added additionally resulting in narrow spatial bandwidth for the anticorrelation[28]. Thus, the violation of $g^{(2)}(\tau = 0) < 1/2$ for the seemingly incoherence-optics-based nondegenerate SPDC-HOM experiments is well explained in Fig. 2 with a simple (quantum) beat model for coherence optics[27,28].

As analyzed above, the anti-correlation or photon bunching phenomenon for $g^{(2)}(\tau=0)=0$ is caused by destructive quantum interference between two input fields at a particular phase, where each output field is a superposition state of the inputs. Furthermore, the destructive quantum interference induces the missing parameter $\varphi_n$ of equation (7). The anticorrelation in HOM experiments has also been explained by wide bandwidth photon pairs in Fig. 2 satisfying equation (6). Understanding the input photons' phase relation satisfying equation (7) is much more important than coincidence timing at detectors, where the intensity correlation of $g^{(2)}$ oscillates as shown in Figs. 1 and 2d. As experimentally demonstrated[24], the modulation period of $g^{(2)}$ exactly matches with wavelength of the SPDC $\chi^{(2)}$ generated input field, satisfying the presented physics of equation (6). In terms of destructive quantum interference, thus, it is intuitively understood how an additional $\pi/2$−phase shift in equation (7) should work together with another BS induced $\pi/2$−phase shift in equation (1). Equation (7) also supports the phase relation between the down-converted photons in SPDC process, where the $\pi/2$−phase shift is to compensate the pump field-induced $\pi$−phase shift on the excited atoms.

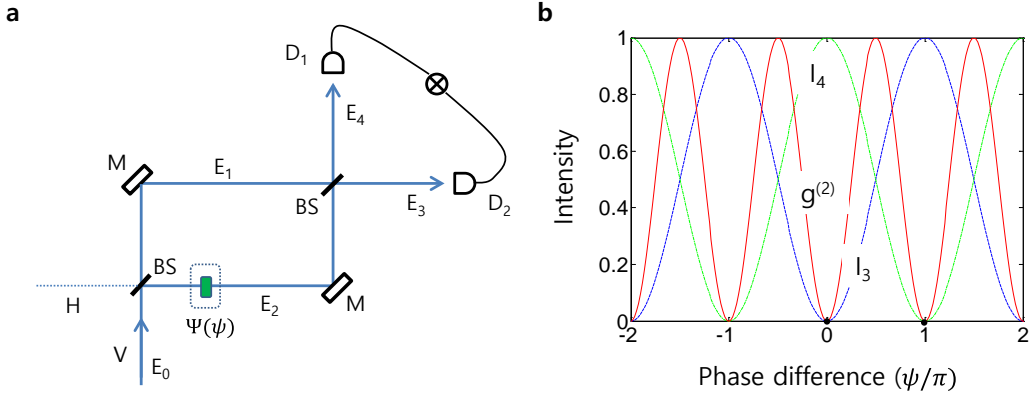

**Figure 3| A Mach-Zehnder interferometer for intensity correlation. a,** A schematic of a coherence-optics-based HOM setup. BS, beam splitter; $E_i$, i$^{th}$ coherent light field; $D_i$, i$^{th}$ photodetector; M, mirror. **b,** Numerical calculations for **a** at $\tau=0$. Red: $g^{(2)}$ correlation (normalized). The blue (green dash-dot) dashed curve is for $I_3$ ($I_4$). At $\psi=\pm 2n\pi$ the output is bunched into $E_4$, whereas into $E_3$ at $\psi=\pm(2n-1)\pi$.

Now, our interest is in the missing parameter and its symmetric property: $\varphi_n=\pm\left(n-\frac{1}{2}\right)\pi$. If a particular sign of $\varphi_n$, say $\varphi_{+1}=+\frac{\pi}{2}$, is assigned to $E_2$ in Fig. 1a, then the bunched photons go for $E_4$ according to equation (2). This input phase-dependent output determinacy is shown in Fig. 3. The photon bunching on a BS satisfying equation (6) follows coherence optics. Because the input photon phase relation of equation (7) can be satisfied by a BS, the anti-correlation scheme of Fig. 1 can also be implemented in Fig. 3. Figure 3a is a typical MZI scheme used for Young's double-slit experiment for $g^{(1)}$ at one output, where the path length (phase) is controlled by the inserted phase shifter $\Psi$. Thus, the output fields in Fig. 3a are described by:

$$\begin{bmatrix}E_3\\E_4\end{bmatrix}=[BS][\Psi][BS]\begin{bmatrix}E_0\\0\end{bmatrix}=\frac{1}{2}\begin{bmatrix}(1-e^{i\psi}) & i(1+e^{i\psi})\\i(1+e^{i\psi}) & -(1-e^{i\psi})\end{bmatrix}\begin{bmatrix}E_0\\0\end{bmatrix}, \qquad (8)$$

where $[\Psi]$ is a phase shifter matrix: $[\Psi]=\begin{bmatrix}1 & 0\\0 & e^{i\psi}\end{bmatrix}$. According to MZI physics, output determinacy is controlled by either $\psi\in\{0,\pi\}$ or incident channel of $E_0$ (V or H): For details, see the Supplementary information C. Here, the incident channel selection or $\psi\in\{0,\pi\}$ in Fig. 3a is equivalent to choosing the sign of $\varphi_n$ in equation (7), determining which-way information in the output modes (see the swapping between dash-dot green curve and dashed blue curve for neighboring anti-correlations (two dots) in Fig. 3b). The MZI directionality is due to destructive quantum interference, where the anti-correlation between two output fields indicates non-classical results[11]. If the two-photon bunching phenomenon on a BS is nonclassical[11], then Fig. 3 satisfying equation (8) is also nonclassical (see two dots in Fig. 3b), violating Bell's inequality (discussed in Discussion). Thus, the anti-correlation ($g^{(2)} = 0$) on a BS in Fig. 1 can be achieved in the MZI scheme of Fig. 3, whether the input $E_0$ is a single photon or coherent fields. The MZI scheme of Fig. 3a for the anti-correlation is the second discovery of the present study.

In SPDC-HOM experiments, the phase relation of independent input photon pairs in time domain does not matter as analyzed in Fig. 2. Moreover, there is a $\pi/2$−phase shift between the down-converted photons. This fact of phase matched input photon pairs has been overlooked for long time in the quantum optics community, even though the input phase modulation has been observed[6]. Depending on a fixed input phase difference, the bunched output channel is deterministically decided. Random output bunching can also be obtained by a random input choice. As a result, an entanglement superposition can be achieved by combining the output modes via two random input modes in $\varphi_n$ (see two dots in Figs. 1(b) and 3(b))[6]: $|\Xi\rangle = \frac{1}{\sqrt{2}}(|2\rangle_3|0\rangle_4 + |0\rangle_3|2\rangle_4)$ for single photons; $|X\rangle = \frac{1}{\sqrt{2}}(|I_0\rangle_3|0\rangle_4 + |0\rangle_3|I_0\rangle_4)$ for coherent lights. For a fixed input channel $E_0$ in Fig. 3, this is accomplished by the random choice of $\psi \in \{0, \pi\}$. Although the coincidence detection $P'$ does not reveal which-way information of the bunched photons (fields), it is clear that the output mode strictly depends on the input choice of missing parameter $\varphi_n$. Thus, the symmetric sign of $\varphi_n$ plays a key role in entanglement superposition of the output modes satisfying Bell's inequality. This result has never been discussed yet.

By adding the missing parameter $\varphi_n$ to $E_1$ in Fig. 1a, the following representations are achieved for the outputs satisfying anti-correlation from equations (4) and (5):

$$E_3 = \frac{1}{\sqrt{2}}(E_1 + iE_2) = \frac{1}{\sqrt{2}}(\pm iE_0 + iE_0) = \sqrt{2}iE_0 \text{ or } 0, \qquad (9)$$

$$E_4 = \frac{i}{\sqrt{2}}(E_1 - iE_2) = \frac{i}{\sqrt{2}}(\pm iE_0 - iE_0) = 0 \text{ or } \sqrt{2}E_0. \qquad (10)$$

Both outputs in equations (9) and (10) are strongly coupled together via the symmetric anti-correlation bases of equation (7), resulting in nonclassical nature. This entanglement superposition composed of two nonclassical output modes has already been discussed for unitary transformation in secured communications[23], where the same analogy should suffice for $\psi \in \{0, \pi\}$ in Fig. 3. In a brief summary, the origin of the anti-correlation for photon bunching in HOM experiments, satisfying nonclassicality[5,6] or Bell's inequality[3,4], is due to destructive quantum superposition on BS between two input modes, where the missing parameter $\varphi_n$ requires a particular phase relation between two input modes. The Bell's inequality also suffices for the linear superposition of the anticorrelation modes determined by the missing parameter $\varphi_n$ with its symmetric property. As a result, the nonclassical feature can be accomplished coherently in MZI. Furthermore, the Schrödinger's cat may be achieved in Fig. 3 via superposition of the anti-correlation modes in coherence optics.

In summary, the second-order anticorrelation between two output modes on a beam splitter (or Young's double-slit) was studied to understand its physical origin of nonclassicality in photon bunching on a beam splitter. Unlike common understanding limited to pure quantum optics of anti-bunched photon nature governed by sub-Poisson statistics, the nonclassical phenomenon of photon bunching in HOM experiments was due to destructive quantum interference with a particular phase relationship satisfying coherence optics. Furthermore, an equivalent model of the beam splitter-based anti-correlation was presented and discussed for a typical MZI scheme. Finally, an entanglement superposition of the output modes of MZI was briefly discussed for potential applications of unconditionally secured classical cryptography. Miscellaneously the SPDC process was analyzed for a HOM dip, where individual input photon pairs governed by incoherence optics results in the anti-correlation owing to the phase matched photon pairs regardless of their frequency detuning. Moreover, the required $\pi/2$ phase shift between the signal and idler photons for a HOM dip was explained as a consequence in the SPDC process. Thus, the border line between classical and quantum optics may be redrawn by coherence optics. The classicality cannot be limited to the wave nature but to incoherence optics. In conclusion, coherence at a particular phase between two input photons is a necessary condition for the nonclassicality of anticorrelation in a HOM dip as well as Bell's inequality. Thus, the present discovery seemingly conflicting with conventional single-photon-based quantum optics opens a door to new regime of quantum information compatible with coherence optics. The significance of this study is to give us a better understanding of the second-order anticorrelation, where quantum superposition results in not only indistinguishability in the input modes but also no which-way information in the output entangled mode.

Acknowledgements
This work was supported by GIST program, S. Korea.
Author contribution
B.H. solely wrote the manuscript.
Additional information
Supplementary information is available in the online version of the paper.
Competing financial interests
The authors declare no competing financial interests.